\documentclass{webofc}
\usepackage[varg]{txfonts}   
\newcommand{\be}{\begin{eqnarray}}
\newcommand{\ee}{\end{eqnarray}}
\newcommand{\ba}{\begin{array}}
\newcommand{\ea}{\end{array}}

\newcommand{\bi}{\begin{itemize}}
\newcommand{\ei}{\end{itemize}}

\woctitle{6th International Conference on Physics Opportunities at Electron-Ion Collider}
\begin{document}
\title{On the universality of the $J=0$ fixed pole contribution in DVCS}
%
%

\author{D.~M\"{u}ller\inst{1}\fnsep\thanks{\email{dieter.mueller@irb.hr}} \and
        K.~Semenov-Tian-Shansky \inst{2}\fnsep\thanks{\email{cyrstsh@gmail.com}} 
}

\institute{Theoretical Physics Division, Rudjer Bo\v{s}kovi\'{c} Institute, HR-10002 Zagreb, Croatia
\and
           Petersburg Nuclear Physics Institute, Gatchina, 188300, St.Petersburg, Russia
          }

\abstract{%
  S.~Brodsky, F.J.~Llanes-Estrada and A.~Szczepaniak formulated the
  $J=0$
  fixed pole   universality hypothesis for (deeply) virtual Compton scattering.
  We show that in the Bjorken   limit this hypothesis is equivalent to the validity
  of the inverse moment sum rule for the $D$-term form factor.
  However, any supplementary
  $D$-term added to a generalized parton distribution (GPD) results in an additional
  $J=0$
  fixed pole contribution  that violates universality. Unfortunately, one can not provide
  any reliable theoretical argument excluding the existence of such supplementary
  $D$-term. Moreover, the violation of
  $J=0$
  fixed pole
  universality  
  was revealed in field theoretical GPD models.
  Therefore, $J=0$ fixed pole   universality hypothesis remains just an external assumption and
  probably will never be proven theoretically.
}
\maketitle
\section{Introduction}
\label{intro_semenov}
Studies of Compton scattering off a nucleon
$\gamma^{(\ast)}(q_1) +  N(p_1) \to \gamma^{(\ast)}(q_2) + N(p_2)$
with both photons being real
($q_1^2= q_2^2=0$)
or with one
($q_1^2=-Q_1^2$, $q_2^2=0$)
or two
($q_1^2=-Q_1^2$, $q_2^2=-Q_2^2$)
virtual space-like photons allow to probe nucleon structure in low and high energy
regimes. The customary theoretical framework which helps to address Compton scattering in
high energy regime is the Regge theory. Assuming the existence of the leading Regge
trajectory
$\alpha(t)$ ($t \equiv (p_2-p_1)^2$),
the high energy leading asymptotic behavior
$\sim s^{\alpha(t)}$
of the amplitude originates from a moving pole in the plane of complex cross-channel
angular momentum
$J$.
Besides these common moving poles (and/or cuts) there also might exist so-called
{\it fixed pole} singularities
(see {\it e.g.} Chapter~I of Ref.~\cite{Alfaro_red_book}),
that do not slide  with the change of
$t$
and can not be revealed by means of the analytic continuation in
$J$.
A $J=J_0$ fixed pole singularity may arise from a cross channel exchange with a
non-Reggeized (elementary) particle of spin
$J_0$
in the cross channel
(or from a contact interaction term). It appears then as the Kronecker-$\delta$
singularity in the complex $J$-plane.

The discussion on the manifestation of the $J=0$ fixed pole singularity
in Compton scattering, which results in constant contribution in high-energy asymptotic limit,
lasts since the late sixties (see the pioneering paper
\cite{Creutz:1968ds} and
Ref.~\cite{Brodsky:1971zh} as well as references therein). S.~Brodsky, F.~Close and J.~Gunion
identified this contribution at the diagrammatic level. They argue it originates from ``seagull''
diagrams corresponding to local two photon interaction.
More recently, in
\cite{Brodsky:2008qu}
S.~Brodsky, F.~J.~Llanes-Estrada and A.~Szczepaniak emphasized that such
$J=0$
fixed pole contribution in virtual Compton scattering is universal
({\it i.e.} independent of photon virtualities) and can be expressed by the inverse moment
of the corresponding $t$-dependent parton distribution function (PDF).

On the other hand, the $J=0$ fixed pole contribution in deeply virtual Compton scattering (DVCS)
is closely related with the $D$-term form factor,
which arises as the subtraction constant for deeply virtual Compton scattering
amplitude within the partonic picture.
The $D$-term
\cite{Polyakov:1999gs}
was originally introduced to complement the polynomiality condition for GPDs within
the double distribution (DD) representation. Later it was realized that it could be implemented as an inherent
part of GPD both within the modified DD representation and within the representations
based on conformal partial wave (PW) expansion of GPDs.

Below, following Refs.~\cite{Muller:2014wxa,Muller:2015vha},
we review the relation between the
$J=0$
fixed pole contribution and the $D$-term form factor.
We tie together the $J$-analytic properties of cross channel
SO$(3)$ PWs and those of GPD conformal moments in the  conformal spin $j$.
We show that the
$J=0$
fixed pole universality hypothesis generally remains unproven and only can
be accepted as external principle for building up phenomenological GPD models.

\section{Dispersive approach for Compton amplitude}

We focus on the Compton form-factor (CFF)
${\cal H}(\nu \! \equiv \! \frac{s-u}{4},t|Q_1^2,Q_2^2)$
occurring in the parametrization of the transverse non-flip
photon helicity amplitude of Compton scattering.  This CFF has even signature ($P=+1$; $C=+1$)
and is the analogue of the usual Dirac electromagnetic form factor. In the forward kinematics the
imaginary part  of
${\cal H}$
corresponds to the deep inelastic structure function
$F_1$.

\begin{figure*}
\centering
\includegraphics[width=5cm,clip]{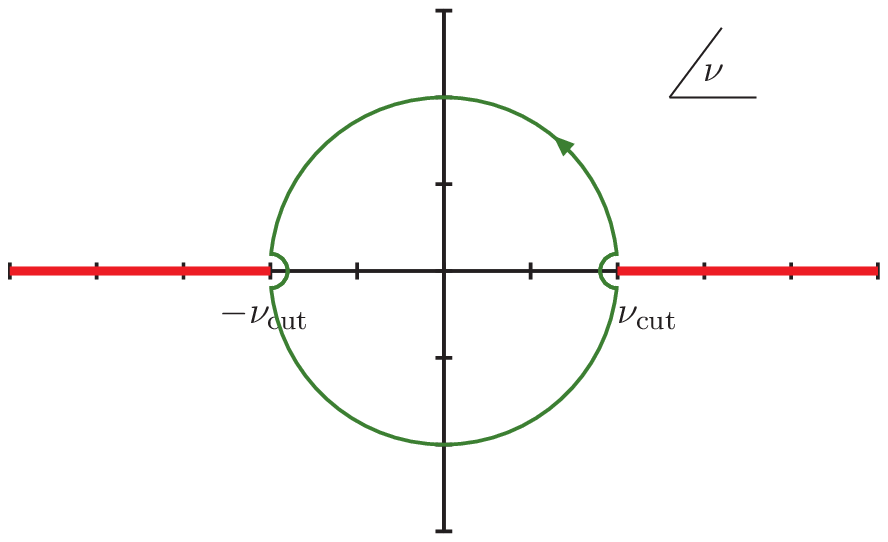} \ \ \ \
\includegraphics[width=4cm,clip]{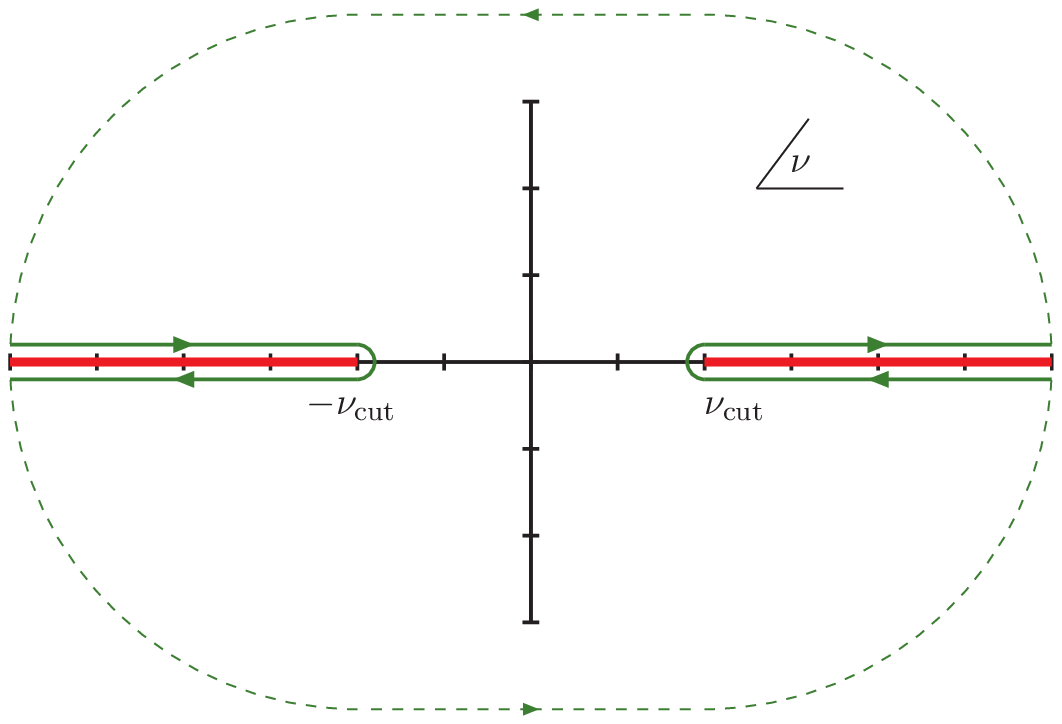}
\caption{Left panel: Initial integration contour in the complex $\nu$ plane to
express the CFF through the Cauchy integral.
 Right panel: Deformation of the integration contour in eq.~(\ref{UnsubtrDR}).}
\label{fig-1_Semenov}       
\end{figure*}

A fixed-$t$ dispersion relation for the CFF ${\cal H}$
can be obtained adopting usual assumptions on the analytic
properties of CFFs. Neglecting the Born term (which is irrelevant in the Bjorken limit)
we consider the deformed integration  contour surrounding cuts on the real axis starting
at the pion production threshold
$\nu_{\rm cut}=\frac{Q_1^2+Q_2^2+t+(M+2m_\pi)^2-M^2}{4 M}$
(see Fig.~\ref{fig-1_Semenov}).
Assuming that
${\cal H}$
vanishes at infinity
($\lim_{|\nu|\to \infty} {\cal H}(\nu,t|Q_1^2,Q_2^2)= 0$), we obtain the unsubtracted DR in the standard form,
\begin{equation}
{\cal H}(\nu,t|Q_1^2,Q_2^2)=
\frac{1}{\pi} \int^{\infty}_{\nu_{\rm cut}} d\nu^\prime \frac{2\nu^\prime\, {\rm Im} \, {\cal H}(\nu^\prime,t|Q_1^2,Q_2^2) }{\nu^{\prime 2} -\nu^2-i \epsilon}.
\label{UnsubtrDR}
\end{equation}
Once ${\cal H}$ does not vanish at infinity one still can formally consider
the unsubtracted DR by specifying a regularization procedure both for the
divergent integral along the real axis and (either divergent) contributions
from large semicircles of the contour on  right panel of Fig.~\ref{fig-1_Semenov}.
A possible choice is the analytic regularization (see {\it e.g.} Ref.~\cite{GelShi64}).
The dispersive integral is replaced by the loop integral in the complex plane that includes the point
$\nu=\infty$, denoted as $(\infty)$.
The unsubtracted DR for CFF
${\cal H}$
then reads
\begin{equation}
\label{DR-anareg-dv}
{\cal H}(\nu,t|Q_1^2,Q_2^2) = {\cal H}_\infty(t|Q_1^2,Q_2^2)
 + \frac{1}{\pi} \int^{(\infty)}_{\nu_{\rm cut}} d\nu^\prime \frac{2\nu^\prime\, {\rm Im} \, {\cal H}(\nu^\prime,t|Q_1^2,Q_2^2)}{\nu^{\prime 2}-\nu^2-i \epsilon},
\end{equation}
where the constant
${\cal H}_\infty$,
arises from the analytic regularization at
$\nu=\infty$.
Within the Regge--pole expansion of the amplitude it is interpreted as the
$J=0$
fixed pole contribution.

An alternative form of DR employed within the deeply virtual  (d.v.) regime
involves one subtraction at the unphysical  point
$\nu=0$:
\begin{equation}
\label{DR-sub-dv}
{\cal H}(\nu,t|Q_1^2,Q_2^2)   \stackrel{\rm d.v.}{=}     {\cal H}_0(t|Q_1^2,Q_2^2)
+
 \frac{1}{\pi} \int^\infty_{\nu_{\rm cut}} \frac{d\nu^\prime}{\nu^\prime}
 \frac{2\nu^2\, {\rm Im} \, {\cal H}(\nu^\prime,t|Q_1^2,Q_2^2)}{\nu^{\prime 2} -\nu^2-i \epsilon} .
\end{equation}
Generally, the subtraction constant
${\cal H}_0(t|Q_1^2,Q_2^2)$
represents an unknown quantity. However, in the deeply virtual regime one can
formulate the external principle allowing to fix the subtraction constant
from the absorptive part of the amplitude. Relying on the operator product
expansion one can check that for equal photon virtualities $Q_1^2=Q_2^2={\cal Q}^2$
(DIS kinematics) due to the current conservation the subtraction constant vanishes
to the leading twist accuracy. Within the kinematics of DVCS the subtraction constant
in
(\ref{DR-sub-dv})
corresponds to the $D$-term form factor
${\cal H}_0(t|Q_1^2={\cal Q}^2,Q_2^2=0)\stackrel{\rm d.v.}{=} 4{\cal D}(t)$.

The dispersion relations (\ref{DR-anareg-dv}) and (\ref{DR-sub-dv})
are supposed to represent the same function. Therefore, the $J=0$
fixed pole contribution
${\cal H}_\infty$
could be related to the subtraction constant
${\cal H}_0$.
Plugging in the algebraic decomposition
$
 \frac{2\nu^\prime}{\nu^{\prime 2} -\nu^2-i \epsilon}=   \frac{1}{\nu^\prime} \frac{2\nu^2}{\nu^{\prime 2} -\nu^2-i \epsilon}  + \frac{2}{\nu^\prime}
$
of the Cauchy kernel  into (\ref{DR-anareg-dv}) and comparing it with (\ref{DR-sub-dv}), we read off the  sum rule
\begin{equation}
\label{sumrule-J=0}
{\cal H}_\infty(t|Q_1^2,Q_2^2) = {\cal H}_0(t|Q_1^2,Q_2^2)
-\frac{2}{\pi} \int^{(\infty)}_{\nu_{\rm cut}}  \frac{d\nu}{\nu}\, {\rm Im}{\cal H}(\nu,t|Q_1^2,Q_2^2)\,,
\nonumber
\end{equation}
expressing the $J=0$ fixed pole contribution through the subtraction constant and the analytically
regularized inverse moment of the absorptive part of the amplitude.

In the deeply virtual kinematics regime the convenient
kinematical variable are the Bjorken-like variable
$
\xi=\frac{Q^2}{(p_1+p_2) \cdot \frac{1}{2}(q_1+q_2)}
$,
where $Q^2=-\frac{(q_1+q_2)^2}{4}$;
and the photon asymmetry parameter%
\footnote{Various definitions of $\vartheta$ encountered in literature differ by
${\cal O}(1/Q^2)$
terms vanishing in the generalized Bjorken limit. }
$
\vartheta = \frac{q_1^2-q_2^2}{q_1^2+q_2^2}+{\cal O}(t/Q^2)
$.
Note that for
$t=0$
$\vartheta=0$
one recovers the usual DIS kinematics, while
$\vartheta=1$
corresponds to the DVCS set up.

Within these scaling variables the analytically regularized DR
(\ref{DR-anareg-dv})
and  the subtracted one
(\ref{DR-sub-dv})
read as follows:
\begin{equation}
{\cal H}(\xi,t|\vartheta) =  \frac{1}{\pi} \int^{1}_{(0)}
\frac{d\xi^\prime}{\xi^\prime} \frac{2\xi^2\, {\rm Im}{\cal H}(\xi^\prime,t|\vartheta)}{\xi^2-\xi^{\prime 2}
-i \epsilon}
+  {\cal H}_\infty(t|\vartheta);
\label{DR-reg1}
\end{equation}
\begin{equation}
{\cal H}(\xi,t|\vartheta) =\frac{1}{\pi} \int^{1}_{0}
d\xi^\prime\frac{2\xi^\prime\, {\rm Im}{\cal H}(\xi^\prime,t|\vartheta)}{\xi^2-\xi^{\prime 2} -i \epsilon}
+  {\cal H}_0(t|\vartheta)\,.
\label{DR-sub1}
\end{equation}
Here, the lower integration limit,
$\xi=0$,
corresponds to
$\nu=\infty$ and
the upper integration limit
$\xi_{\rm cut} =\frac{Q^2}{2 M\nu_{\rm cut}}$,
has been set
to $\xi_{\rm cut}=1$
in the (generalized) Bjorken limit.
The equivalence of the two DRs
(\ref{DR-reg1}),
(\ref{DR-sub1})
is ensured by the sum rule
(\ref{sumrule-J=0}),
which  now reads
\begin{equation}
{\cal H}_\infty(t|\vartheta) =  {\cal H}_0(t|\vartheta) -
\frac{2}{\pi} \int^{1}_{(0)} \frac{d\xi}{\xi}\,  {\rm Im}{\cal H}(\xi,t|\vartheta)\,.
\label{sum-rule-1}
\end{equation}
Within these notations, the
$J=0$
fixed pole universality hypothesis of
\cite{Brodsky:2008qu}
reads:
\begin{equation}
\label{J=0-conjecture}
{\cal H}_\infty(t|\vartheta)=
-\frac{2}{\pi} \int^{1}_{(0)} \frac{d\xi}{\xi}\, {\rm Im}{\cal H}(\xi,t|\vartheta=0)\,.
\end{equation}

\section{Fixed pole contribution in GPD framework}

In this section we address the issue of the subtraction constant
for the dispersive representation of CFF within the partonic picture.
We would now like to point out the origin of additional fixed
pole contribution
$\Delta {\cal H}_\infty$,
which violates the universality conjecture within the GPD framework.

Within the collinear factorization approach to the leading order (LO) accuracy, the CFF
${\cal H} (\xi,t|\vartheta)$
arises from the elementary amplitude
\begin{equation}
{\cal H} (\xi,t|\vartheta) \stackrel{\rm LO}{=}
  \int_{0}^1\!dx\,\frac{2x}{\xi^2-x^2- i \epsilon}
  H^{(+)} (x,\eta=\vartheta\xi,t)\,,
\label{eq:CFFLO}
\end{equation}
where
$H^{(+)}(x,\eta,t)\equiv   H(x,\eta,t) - H(-x,\eta,t)$ 
stands for the antisymmetric charge even quark GPD combination.
Note that for all allowed values of
$|\vartheta| \le 1$
the imaginary part of the CFF is given by the GPD in the outer region
$\xi \ge \eta$.

Inserting the imaginary part of (\ref{eq:CFFLO})
into the sum rule (\ref{sum-rule-1})
allows to express the $J=0$ fixed pole contribution
${\cal H}_\infty(t|\vartheta)$ to the LO accuracy by
the GPD in the outer region:
\begin{equation}
{\cal H}_\infty(t|\vartheta) \stackrel{\rm LO}{=}
{\cal H}_0(t|\vartheta)
- 2 \int^{1}_{(0)} \frac{dx}{x}\, H^{(+)}(x,\vartheta x,t)\,.
\label{sum-rule-GPD}
\end{equation}
By plugging the imaginary part of (\ref{eq:CFFLO})
into the subtracted DR (\ref{DR-sub1})
and equating it with the LO convolution formula (\ref{eq:CFFLO}) for the CFF,
one can work out the GPD sum rule \cite{Anikin:2007yh,Kumericki:2008di}:
\begin{equation}
{\cal H}_0(t|\vartheta) \equiv 4{\cal D}(t|\vartheta) \stackrel{\rm LO}{=}\int_{0}^1\!dx \frac{2x}{x^2-\xi^2}
 \left[\! H^{(+)} (x,\vartheta x,t) -H^{(+)} (x,\vartheta \xi,t)\right]
\label{D-sumrule}
\end{equation}
for the $D$-term form factor, which corresponds to the subtraction constant
${\cal H}_0$.
It seems that by formally taking the high energy limit
$\xi \to 0$
one can provides a regular proof for the $J=0$ fixed pole universality conjecture
(\ref{J=0-conjecture}).
We argue that one has to be prudent while performing this limiting procedure and
separate the integration region in
(\ref{sum-rule-GPD})
into central
$x \in [0; \theta \xi]$
and outer
$x \in [\theta \xi; 1]$
regions. Omitting the contribution from the central region means ignoring
possible separate $D$-term addenda to GPD. As an illustration, we considered
the specific form of a DD representation for GPD:
\begin{equation}
H^{(+)}(x,\eta) =
\int_{0}^{1}\!dy\int_{-1+y}^{1-y}\!dz\, \Big[(1-a x)\delta(x-y-z\eta)
 - \{x\to -x\} \Big] h(y,z)\,,
\label{DD2GPD}
\end{equation}
where $h(y,z)$ is DD, symmetric in $z$ and antisymmetric in $y$.
The case $a=0$ corresponds to the usual DD representation (without a $D$-term).
For $a \ne 0$ GPD
(\ref{DD2GPD})
includes an inherent $D$-term part and the polynomiality
condition is respected in its complete form. The GPD sum rule
(\ref{D-sumrule})
is also respected and the
$J=0$
fixed pole universality conjecture
(\ref{J=0-conjecture})
is valid.

However, adding a supplementary separate $D$-term contribution
\begin{equation}
H^{(+)}(x,\eta) \to H^{(+)}(x,\eta) + \theta(|x|\le |\eta|) d^{\rm f.p.}(x/|\eta|)
\label{DD+D_representation}
\end{equation}
leads to breakdown of the
$J=0$
fixed pole universality conjecture by the
$J=0$
fixed pole contribution
$4 {\cal D}^{\rm f.p.}(t|\vartheta)=\int_{0}^{1}\!dx\, \frac{2 x \vartheta^2 }{1-\vartheta^2 x^2 }\,
d^{\rm f.p.}(x,t)$:
\begin{equation}
\label{H_infty-sum_rule}
{\cal H}_\infty(t|\vartheta) \stackrel{\rm LO}{=}
-2 \int^{1}_{0} \frac{dx}{x}\, q^{(+)}(x,t)+\underbrace{4 {\cal D}^{\rm f.p.}
(t|\vartheta)}_{\Delta{\cal H}_\infty(t|\vartheta)}\,.
\end{equation}

A particular example of a field theoretical GPD model with a separate
$D$-term contribution is provided by the calculation
\cite{SemenovTianShansky:2008mp}
of pion GPDs in the non-local chiral quark model
\cite{Praszalowicz:2003pr}.
In this model the universality conjecture
(\ref{J=0-conjecture})
is not valid due to a supplementary
$D$-term
$d^{\rm f.p.}(x,t)$
contribution, which has to be added to make GPD satisfy the soft pion theorem
\cite{Pobylitsa:2001cz}
fixing pion GPDs in the limit
$\eta \to 1$.
Another example is given by the photon GPD $
H_1$, calculated to one-loop accuracy in
\cite{Friot:2006mm},
where a direct calculation of the CFF in the
$\xi \to 0$
limit disproves the conjecture
(\ref{J=0-conjecture}).

We conclude that the validity of the
$J=0$
fixed pole universality conjecture corresponds  to the
internal duality principle for GPDs, relating GPDs in the inner
and outer support regions
(see Ref.~\cite{Kumericki:2008di}
for a detailed discussion).

It is extremely convenient to address this property within the approach
\cite{Kumericki:2007sa}
based on operator product expansion over the conformal basis. In this case
there turns to be two kinds of analytical properties relevant for GPDs
and associated CFFs:
\begin{itemize}
\item  analyticity of CFFs in the cross channel angular momentum $J$;
\item  analyticity of GPD Gegenbauer/Mellin moments in the variable $j$, labeling the conformal spin $j+2$
of twist-$2$ quark conformal basis operators %
\footnote{Here the covariant derivative $\stackrel{\leftrightarrow}{D}$
and the total derivative $\stackrel{\leftrightarrow}{\partial}$ are contracted with the light-cone vector $n$,
 $C_j^{\frac{3}{2}}$
 stand for the usual Gegenbauer polynomials with the index $3/2$. }
${\cal O}_j^a=
\frac{\Gamma(3/2)\Gamma(1+j)}{2^{j} \Gamma(3/2+j)}\, (i \!
\stackrel{\leftrightarrow}{\partial}_+)^j\; \bar{\psi} \lambda^a \gamma_+ \, C_j^{3/2}\!\left(\!
\frac{\stackrel{\leftrightarrow}{D}_{+}}{\stackrel{\leftrightarrow}{\partial}_+}\!
\right)\psi.$
\end{itemize}

In order to show explicitly the $J$-analytic properties we
employ the Froissart-Gribov projection
\cite{Gribov1961}
of the cross channel
${\rm SO}(3)$ PWs
of the CFF:
$
a_{J}
(t|\vartheta) \equiv \frac{1}{2}\int_{-1}^{1}\!d(\cos\theta_t)\,
P_J(\cos\theta_t)  {\cal H}^{(+)}(\cos \theta_t,t|\vartheta),
$
where (neglecting the threshold corrections) the cosine of the $t$-channel
scattering angle reads:
$\cos \theta_t= -1/(\vartheta \xi)+ {\cal O}(1/Q^2)$.

For $J>0$ PWs the Froissart-Gribov projection provides
to the LO accuracy
\begin{equation}
a_{J>0}(t|\vartheta) \stackrel{\rm LO}{=}
2 \int_0^1\! dx
\frac{{\cal Q}_J(1/x)}{x^2} H^{(+)}(x,\vartheta x,t)\,,
\label{FroissartGribov}
\end{equation}
where ${\cal Q}_J(1/x)$ denote the Legendre functions of the second kind.
For $J=0$ one obtains
\begin{equation}
a_{J=0}(t|\vartheta)  \stackrel{\rm LO}{=}   2 \int_0^1\! dx \left[\frac{{\cal Q}_0(1/x)}{x^2} -
\frac{1}{x}\right] H^{(+)}(x,\vartheta x,t)
+ 4 {\cal D}(t|\vartheta)\,.
\label{FroissartGribovJ0}
\end{equation}

Indeed, as clearly seen from Eqs.~(\ref{FroissartGribov}) and
(\ref{FroissartGribovJ0}),
one can not recover $a_{J=0}(t|\vartheta)$
by means of the analytic continuation of
$a_{J>0}(t|\vartheta)$ to $J=0$.
Therefore, analyticity in the cross channel angular momentum $J$
is affected by the presence of a $J=0$ fixed pole contribution
\begin{equation}
a_{J=0}^{\rm f.p.}(t|\vartheta)
\equiv {\cal H}_\infty(t|\vartheta)
\stackrel{\rm LO}{=}  4 {\cal D}(t|\vartheta) -
2 \int^{1}_{(0)} \frac{d x}{x}\,  H^{(+)}(x,\vartheta x,t)\,.
\label{a_{J=0}^{f.p.}}
\end{equation}
Within the conformal operator product expansion approach
the analytic properties in the conformal spin $j$ of the Gegenbauer moments of GPDs
play the crucial role. In particular, the presence of a $J=0$ fixed pole contribution
(\ref{a_{J=0}^{f.p.}})
to the CFF ${\cal H}$ is a direct consequence of
absence of non-singular conformal operators with the Lorentz spin
$J\equiv j+1=0$.
Such a
$j=-1$
contribution has to be subtracted from the
$J=0$
PW (see the $1/x$ moment in
the integrand of Eq.
(\ref{FroissartGribovJ0})).
The analogous cancelation appears also in the framework of dual parametrization of GPDs
\cite{Muller:2014wxa}.
Moreover, assuming maximal analyticity in the conformal spin
$j$
(which corresponds to absence of the
$j=-1$
fixed pole singularities manifest as the Kronecker
$\delta_{j, \,-1}$
terms
or singular operator contributions in the momentum fraction representation) leads
to the validity of the internal duality principle for GPDs
(see \cite{Kumericki:2008di}).
This corresponds the implementation of $J=0$ fixed pole universality hypothesis with a definite
$J=0$
fixed pole  contribution given by the inverse moment sum rule
\begin{equation}
{\cal H}_\infty(t|\vartheta)  \stackrel{\rm LO}{=}
-2 \int^{1}_{(0)} \frac{dx}{x}\, H^{(+)}(x,0,t).
\label{InvPdf}
\end{equation}
It is worth emphasizing that the inverse moment of PDF
(\ref{InvPdf})
can not be extracted form the sum rule
(\ref{D-sumrule})
for $D$-term form factor as it simply cancels out.
This sum rule is only potentially sensitive to possible non-universal
contribution into
${\cal H}_\infty(t|\vartheta)$.

 \section*{Conclusions}
The $J=0$ fixed pole universality hypothesis remains an attractive theoretical assumption for
building up consistent GPD models intended for description of Compton scattering in
deeply virtual regime. However, at the moment this hypothesis lacks theoretical proof.
Phenomenological verification of this conjecture from the GPD sum rule
(\ref{sum-rule-GPD})
is in principle possible but is biased by the necessity of extrapolation of experimental
results into high energy asymptotic regime and therefore will hardly be considered as decisive.

\end{document}